\documentclass[aps,prl,reprint,preprintnumbers,superscriptaddress,amsmath,amssymb,bibnotes,longbibliography]{revtex4-1}

\usepackage{graphicx}
\usepackage{dcolumn}
\usepackage{bm}
\usepackage[colorlinks,linkcolor=blue,anchorcolor=blue,citecolor=blue,urlcolor=blue,filecolor=blue,menucolor=blue,runcolor=blue]{hyperref}

\begin{document}
\title{Structural and magnetic properties of antiferromagnetic Ce$_2$IrGa$_{12}$}
\author{Y. J. Zhang}
\affiliation{Center for Correlated Matter and Department of Physics,
Zhejiang University, Hangzhou 310058, China}
\author{B. Shen}
\affiliation{Center for Correlated Matter and Department of Physics,
Zhejiang University, Hangzhou 310058, China}
\author{F. Du}
\affiliation{Center for Correlated Matter and Department of Physics,
Zhejiang University, Hangzhou 310058, China}
\author{Y. Chen}
\affiliation{Center for Correlated Matter and Department of Physics,
Zhejiang University, Hangzhou 310058, China}
\author{J. Y. Liu}
\affiliation{Department of Chemistry, Zhejiang University, Hangzhou 310027, China}
\author{Hanoh Lee}
\email{hlee.rtf@gmail.com}
\affiliation{Center for Correlated Matter and Department of Physics,
Zhejiang University, Hangzhou 310058, China}
\affiliation{Center for Quantum Materials and Superconductivity, Department of Physics,
SungKyunKwan University, Suwon 16419, Republic of Korea}
\author{M. Smidman}
\email{msmidman@zju.edu.cn}
\affiliation{Center for Correlated Matter and Department of Physics,
Zhejiang University, Hangzhou 310058, China}
\author{H. Q. Yuan}
\affiliation{Center for Correlated Matter and Department of Physics,
Zhejiang University, Hangzhou 310058, China}
\affiliation{Collaborative Innovation Center of Advanced
Microstructures, Nanjing University, Nanjing 210093, China}
\date{\today}
\addcontentsline{toc}{chapter}{Abstract}

\begin{abstract}

 We report a study of the structural and magnetic properties of single crystals of Ce$_2$IrGa$_{12}$. Ce$_2$IrGa$_{12}$ crystallizes in a layered tetragonal structure, and undergoes an antiferromagnetic transition below 3.1~K. We characterize the temperature-field phase diagrams of Ce$_2$IrGa$_{12}$ for fields both within the $ab$-plane and along the $c$-axis, where the presence of a field-induced magnetic phase is found for in-plane fields. The ordering temperature is moderately enhanced upon the application of pressures up to 2.3~GPa, suggesting that Ce$_2$IrGa$_{12}$ corresponds to the well localized region of the Doniach phase diagram.
\begin{description}
\item[PACS number(s)]
 74.70.-b, 71.45.Lr, 74.10.+v
\end{description}
\end{abstract}

\maketitle

\section{Introduction}
Ce-based intermetallic compounds have attracted extensive interest owing to their exotic properties and fascinating underlying physics, such as complex magnetic order, unconventional superconductivity, and quantum criticality \cite{Weng_2016, Pfleiderer_2009, Si_2010}. In these systems, the periodically arranged localized $4f$ moments are screened by the conduction electrons via the Kondo effect. Another competing interaction is the Ruderman-Kittel-Kasuya-Yosida (RKKY) interaction which favors long-range magnetic order \cite{Ruderman_1954, Kasuya_1956, Yosida_1957}, and the competition between these interactions gives rise to a variety of ground states \cite{Doniach_1977}, which may be tuned by pressure, magnetic fields or chemical doping \cite{Weng_2016}.

Ce$_n$$M$In$_{3n+2}$ ($M$=transition metal) systems show a variety of unusual phenomena, including the coexistence of magnetism and superconductivity, non-Fermi liquid behavior near a magnetic instability and multiple quantum phase transitions upon tuning with different parameters \cite{Kim_2001, Tuson_2006, Jiao_2015}. Ce$_n$$M$In$_{3n+2}$ are formed from $M$In$_2$ and CeIn$_3$ layers, which are stacked along the $c$-axis \cite{Thompson_2012}, where CeIn$_3$ itself exhibits antiferromagnetic order below $T_N$=10.1~K at ambient pressure, as well as a pressure induced superconducting dome with a maximum  transition temperature of $T_c^{max}$=0.2~K near a quantum critical point \cite{Knebel_2002}. Among the Ce$M$In$_5$ series, CeCoIn$_5$ \cite{Petrovic_2001} and CeIrIn$_5$ \cite{Petrovic_2001_CeIrIn5} are unconventional superconductors at ambient pressure with $T_c$=2.3~K, and 0.4~K respectively, while CeRhIn$_5$ shows antiferromagnetic ordering below $T_N$=3.8~K, and pressure-induced superconductivity with $T_{\rm c}$ reaching 2.1~K at 2.1~GPa \cite{Hegger_2000}. These findings gave a strong indication that increasing the quasi-two-dimensionality can significantly enhance the $T_c$ of heavy fermion superconductors.

In comparison, the  Ce$_2M$In$_8$ compounds consist of two CeIn$_3$ layers separated by a layer of $M$In$_2$, and therefore the dimensionality is between that of CeIn$_3$ and Ce$M$In$_5$ \cite{Cornelius_2001}. Correspondingly, Ce$_2$CoIn$_8$ \cite{Chen_2002} and Ce$_2$PdIn$_8$ \cite{Kaczorowski_2009} are heavy fermion superconductors at ambient pressure with intermediate $T_c$ values of 0.4~K and 0.68~K, respectively, while Ce$_2$RhIn$_8$ is a heavy fermion antiferromagnet exhibiting pressure induced superconductivity with a maximum $T_c$ of 2~K near 2.3~GPa \cite{Nicklas_2003}. Interestingly, Ce$_2$PdIn$_8$ was found to exhibit nodal superconductivity together with a field-induced quantum critical point at the upper critical field \cite{Dong_2011}, which is strikingly similar to the behavior observed in CeCoIn$_5$ \cite{Petrovic_2001,Paglione_2003}. Meanwhile, the crystal structure of Ce$_3$PtIn$_{11}$ and Ce$_3$PdIn$_{11}$ has two-inequivalent Ce sites, and both compounds show the coexistence at ambient pressure of complex magnetic ground states with multiple magnetic transitions, and heavy fermion superconductivity with $T_c$ of 0.3~K and 0.4~K, respectively \cite{Ce3PtIn11_2015,Das_2018,Ce3PdIn11_2015, Das_2019}.

Efforts have been made to look for isostructural  gallides in the Ce$_nM$Ga$_{3n+2}$ series with Ga instead of In. This is in particular due to the discovery that PuCoGa$_5$ is a superconductor with $T_c=18.5$ K \cite{Sarrao_2002}, which is the largest among the heavy fermion superconductors, and much higher than isostructural PuCoIn$_5$ \cite{Bauer_2011, Ramshaw_2015}. Instead however, a number of different structures are found to occur. CeGa$_6$ has in a tetragonal structure (space group $P4/nbm$), and is an antiferromagnet with $T_N$$\sim$1.7~K \cite{Erik_1999}. Ce$_2$$M$Ga$_{12}$ ($M$= Cu, Ni, Pt, Pd, Rh) crystallize in a layered tetragonal structure with the same space group $P4/nbm$, in which layers of CeGa$_6$ are inserted between two layers of $M$Ga$_6$. Ce$_2$PdGa$_{12}$ is a strongly anisotropic antiferromagnet with $T_N$=11~K, where a ferromagnetic in-plane component appears below a second transition at 5~K \cite{Macaluso_2005, Gnida_2013}.  Ce$_2$NiGa$_{12}$ and Ce$_2$RhGa$_{12}$ are reported to order antiferromagnetically at 10 and 3.5~K, respectively \cite{Cho_2008, Nallamuthu_2014}, while Ce$_2$PtGa$_{12}$ exhibits two antiferromagnetic transitions at $T_{N1}$=7.3~K and $T_{N2}$=5.5~K \cite{Sichevych_2012}. Upon applying pressure, $T_N$ of Ce$_2$PdGa$_{12}$ is suppressed, giving rise to a non-magnetic state at a critical pressure of around $P_c$$\sim$7~GPa \cite{Ohara_2012}. Meanwhile the antiferromagnetism of  Ce$_2$NiGa$_{12}$ disappears at $P_c$$\sim$5.5~GPa while the Ce valence drastically changes at a higher pressure of $P_V$$\sim$9~GPa \cite{Kawamura_2014}.

Compared to the aforementioned two compounds, Ce$_2$RhGa$_{12}$ has much lower $T_N$, and it is of particular interest to examine Ce$_2$$M$Ga$_{12}$ compounds with a lower $T_N$, in order to determine whether they are situated in closer proximity to quantum criticality.
Here we report the synthesis of single crystals of isostructural Ce$_2$IrGa$_{12}$ using a flux method. The physical properties are investigated using electrical resistivity, magnetic susceptibility and specific-heat measurements and we find that it is an antiferromagnet below $T_N $=3.1~K at ambient pressure. The temperature-field phase diagrams are constructed for $H||ab$ and $H||c$, where $T_N$ is suppressed upon applying fields, and a metamagnetic transition is observed for fields applied within the $ab$ plane. $T_N$ shows a moderate enhancement upon applying pressures up to 2.3~GPa, indicating that Ce$_2$IrGa$_{12}$ is situated on the well localized side of the Doniach phase diagram.

\section{Experimental details}

Single crystals of Ce$_2$IrGa$_{12}$ were grown using a flux method. Ce$_2$Ir was first prepared by arc-melting Ce ingot ($99.9\%$) and Ir powder ($99.99\%$) under a titanium-gettered argon atmosphere. Ce$_2$Ir and Ga ($99.99\%$) were then placed in an alumina crucible in a 1:40 atomic ratio and sealed in an evacuated quartz tube. The tube was heated up to 1150$~^\circ$C and held at this temperature for 24 hours, before being cooled slowly to 400$~^\circ$C and centrifuged to remove excess Ga. Shiny cuboid-like samples, with typical lengths of 0.2-0.55~mm were obtained, and the phase was determined to be Ce$_2$IrGa$_{12}$ using single crystal x-ray diffraction (XRD) and a cold field emission scanning electron microscope (SEM) equipped with an x-ray energy spectrometer (EDS). Images of the Ce$_2$IrGa$_{12}$ crystals are shown in Fig.1(b), which were obtained using the SEM. The single crystals of La$_2$IrGa$_{12}$, which were used as a nonmagnetic analog, were synthesized using an analogous method. Single crystal XRD measurements were performed using an Xcalibur, Atlas, Gemini ultra diffractometer with an x-ray wavelength of $\lambda=0.71073$~\AA. The electrical resistivity, magnetic susceptibility and heat capacity were all measured using a Physical Property Measurement System (PPMS-14~T), including a vibrating sample magnetometer option and a $^3$He refrigerator insert. The electrical transport measurements under pressure were carried out in a piston-cylinder clamp-type cell. The resistivity measurements were performed with the current in the $ab$-plane, with the field and current directions perpendicular.

\begin{figure}[h]
\includegraphics[width=8.6cm]{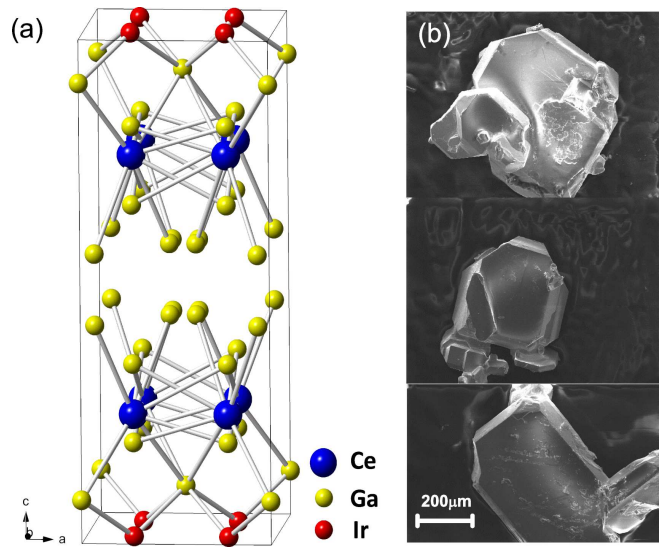}
\caption{(Color online) (a) Crystal structure of Ce$_2$IrGa$_{12}$ where the blue, red and yellow represent the Ce, Ir and Ga atoms, respectively. (b) Scanning electron microscope image  showing the morphology of the Ce$_2$IrGa$_{12}$ single crystals.
}
\label{fig1}
\end{figure}

\section{results}
\subsection{Crystal structure}

\begin{table}[tb]
\caption{Results from refining single crystal XRD measurements of Ce$_2$IrGa$_{12}$, where the refinement parameters, $R_1$, $wR_2$, atomic coordinates and isotropic displacement parameters $U_{eq}$ are displayed.}
\label{ResTab}

\begin{tabular}{p{0.5in} p{0.7in} p{0.7in} p{0.9in} c}
\hline \hline \\[-5pt]
\multicolumn{2}{p{1.2in}}{Formula}&\multicolumn{2}{p{1in}}{Ce$_2$IrGa$_{12}$ }&\\ [5pt]
\multicolumn{2}{p{1.2in}}{Space group}&\multicolumn{3}{p{1in}}{$P4/nbm$~~(No.~125) }\\ [5pt]
\multicolumn{2}{p{1.2in}}{Lattice parameters }&\multicolumn{3}{p{1in}}{$a$=6.0614(4)~\AA,~ $c$=15.6818(13)~\AA}\\ [5pt]
\multicolumn{2}{p{1.2in}}{$R_1$,  $wR_2$ }&\multicolumn{2}{p{1in}}{0.0529,  0.1504 }& \\ [5pt]
\hline \\[-5pt]
Atom & x & y & z & $U_{eq}$\\ [5pt]
Ir1 & 0.75 & 0.25 & 0.00 & 0.0059(5)\\ [5pt]
Ce1 & 0.75 & 0.25 & 0.24261(9) & 0.0063(5)\\ [5pt]
Ga1 & 0.5002(2) & 0.4998(2) & 0.08626(12) & 0.0075(6) \\ [5pt]
Ga2 & 0.25 & 0.25 & 0.1864(2) & 0.0078(7)\\ [5pt]
Ga3 & 0.25 & 0.25 & 0.34204(19) & 0.0117(8)\\ [5pt]
Ga4 & 0.4304(3) & 0.5696(3) & 0.42914(14) & 0.0211(7) \\[5pt]
\hline\hline
\end{tabular}
\end{table}

\begin{figure}[h]
\includegraphics[width=8.6cm]{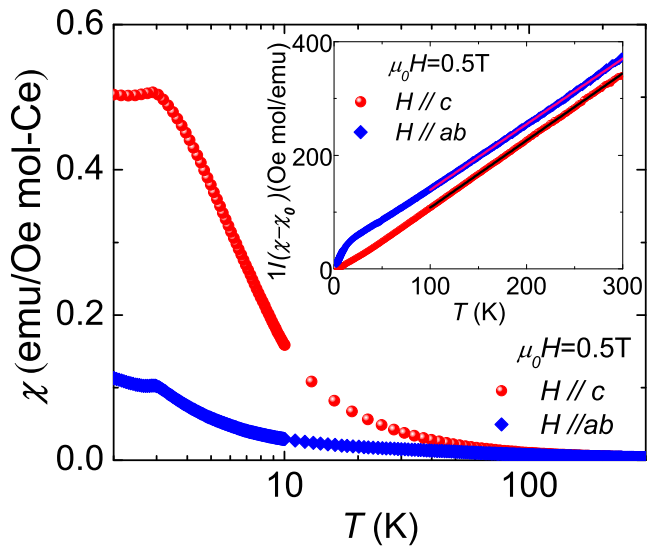}
\caption{(Color online) Temperature dependence of the magnetic susceptibility $\chi(T)$ of Ce$_2$IrGa$_{12}$ measured in an applied field of 0.5~T both parallel to the $c$ axis and in the $ab$ plane. The inset shows 1/($\chi$-$\chi_0$) for the two field directions, and the lines show the Curie-Weiss fitting.
 }
\label{fig2}
\end{figure}

Single crystal x-ray diffraction was performed so as to characterize the crystal structure. The results of the refinement of the data are displayed in Table I. These show that Ce$_2$IrGa$_{12}$ crystallizes in the same tetragonal structure as other Ce$_2$$M$Ga$_{12}$ compounds with space group $P4/nbm$ \cite{Macaluso_2005, Nagalakshmi_2011, Cho_2008}, as displayed in Fig. 1(a). The refined lattice parameters are $a$= 6.0614(4)~\AA, and $c$=15.6818(13)~\AA, which are similar to those of Ce$_ 2$RhGa$_{12}$  \cite{Nagalakshmi_2011}.

\subsection{Antiferromagnetic transitions in Ce$_2$IrGa$_{12}$ }

Figure 2 displays the temperature dependence of the magnetic susceptibility $\chi(T)$ of Ce$_2$IrGa$_{12}$ measured in an applied field of $\mu_0$$H$= 0.5~T both parallel to the $c$ axis and within the $ab$ plane. The low-temperature $\chi(T)$ for $H||c$ and $H||ab$ both exhibit a peak at $T_N$. The lack of hysteresis between zero-field-cooling and field-cooling measurements indicates that the transition is antiferromagnetic. The value of $\chi$ for $H||c$ is higher than for $H||ab$, indicating that the $c$-axis is the easy axis of magnetization. For both field directions, the data of Ce$_2$IrGa$_{12}$ follows Curie-Weiss behavior from 300 K, down to around 100~K. The magnetic susceptibility for both orientations was fitted to a modified Curie-Weiss law: $\chi$=$\chi_0$+$C$/$(T-\theta_P)$ from 100~K to 300~K, where $\chi_0$ is the temperature-independent term, $C$ is the Curie constant and $\theta_P$ is the Curie-Weiss temperature. The derived effective moments are 2.57~$\mu_B$/Ce ($H||c$) and 2.58~$\mu_B$/Ce ($H||ab$), with $\theta_P$=10.2(2)~K ($H||c$) and -22.6(5)~K ($H||ab$). These indicate an anisotropic $\theta_P$, where the spins are antiferromagnetically coupled within the $ab$-plane and ferromagnetically coupled along the $c$-axis.

\begin{figure}[h]
\includegraphics[width=8.6cm]{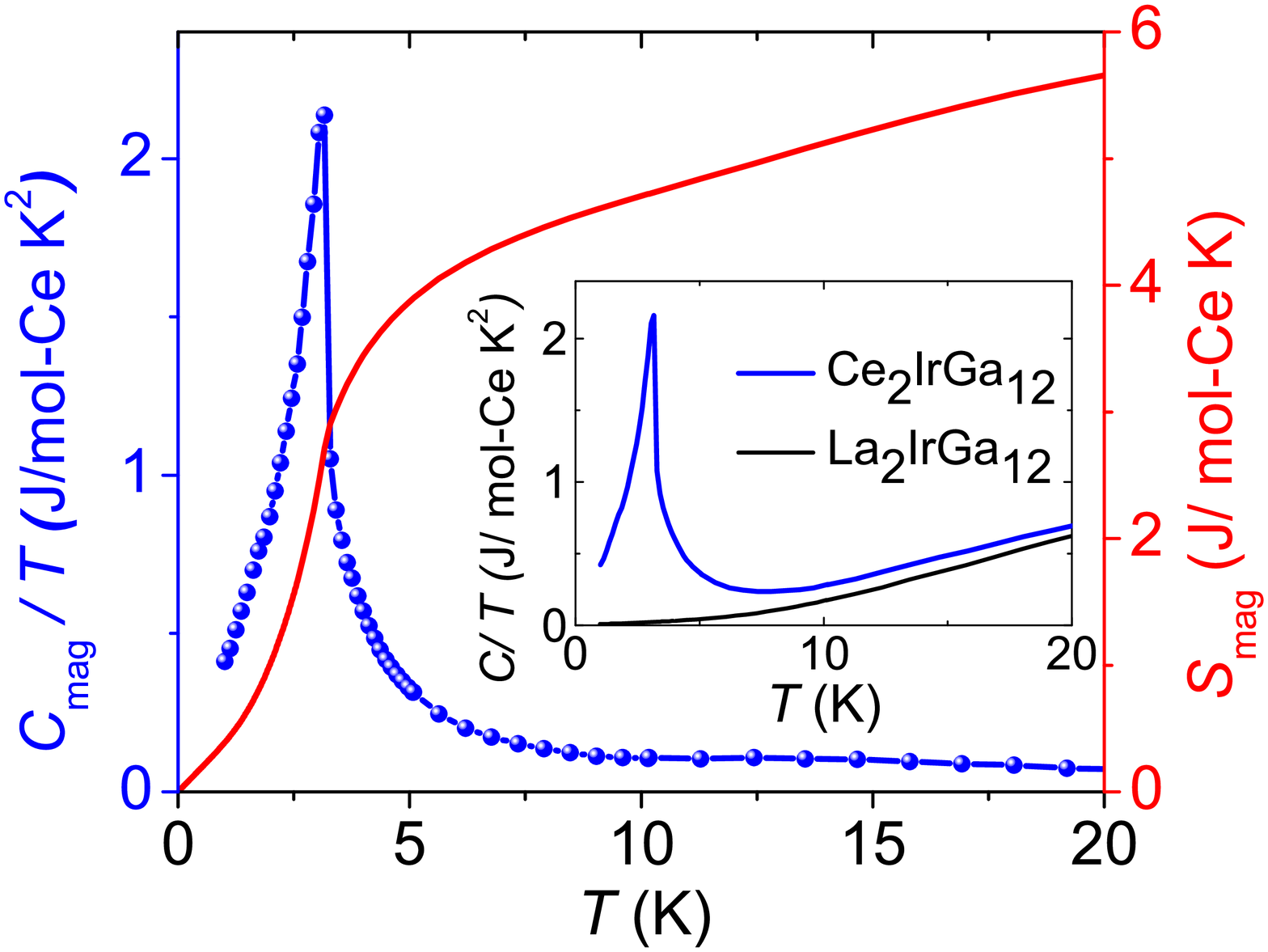}
\caption{(Color online)  Temperature dependence of the magnetic specific heat $C_{mag}/T$ and magnetic entropy $S_{mag}$ for Ce$_2$IrGa$_{12}$, where $C_{mag}/T$ is obtained from subtracting an estimate of the lattice contribution from the La-analog. In order to estimate $S_{mag}$, $C_{mag}/T$ was extrapolated to zero temperature \cite{Continentino_2001}. The inset displays the total specific heat as $C/T$ of Ce$_2$IrGa$_{12}$ (blue line) and La$_2$IrGa$_{12}$ (black line).
 }
\label{fig3}
\end{figure}

Figure 3 shows the temperature dependences  of the magnetic specific heat $C_{mag}/T$ and entropy $S_{mag}$ of Ce$_2$IrGa$_{12}$, where $C_{mag}/T$ was obtained by subtracting the lattice contribution, estimated from the data of nonmagnetic isostructural La$_2$IrGa$_{12}$. The total specific heat of Ce$_2$IrGa$_{12}$ and La$_2$IrGa$_{12}$ are displayed in the inset. A sharp jump in $C_{mag}/T$ occurs at $T_N$=3.1~K, which is typical of a second order phase transition. The specific heat of La$_2$IrGa$_{12}$ below 5~K was fitted using $C/T$=$\gamma$+$\beta$$T^2$, giving rise to a Sommerfeld coefficient $\gamma$=16.6(6)~mJ/mol~K$^2$ and $\beta$=2.84(4)~mJ/mol~K$^4$. A Debye temperature of $\theta_D$=217.4~K was calculated using $\theta_D$=$\sqrt[3]{12\pi^4nR/5\beta}$, where $n$=15 is the number of atoms per formula unit, and $R$=8.314J/mol~K. The same expression was fitted to the data of Ce$_2$IrGa$_{12}$ in the paramagnetic state between 10 and 20~K, giving rise to $\gamma$= 144.4~mJ/mol-Ce~K$^2$, $\beta$=2.86(8)~mJ/mol~K$^4$, and $\theta_D$=216.8~K. A similarly large $\gamma$ of 140~mJ/mol-Ce~K$^2$ was reported for Ce$_2$PdGa$_{12}$ obtained from data at $T$$>$$T_N$ \cite{Macaluso_2005}, while larger values of 191~mJ/mol-Ce~K$^2$ and 212~mJ/mol-Ce~K$^2$ were found for Ce$_2$NiGa$_{12}$, and Ce$_2$RhGa$_{12}$ \cite{Cho_2008,Nallamuthu_2014}. However we note that an enhanced $\gamma$ deduced from data above $T_N$ does not necessarily indicate heavy fermion behavior with large effective carrier masses, but may arise due to the presence of magnetic correlations above $T_N$. The magnetic entropy $S_{mag}$ of Ce$_2$IrGa$_{12}$ was obtained by integrating $C_{mag}/T$ of Ce$_2$IrGa$_{12}$. The magnetic entropy $S_{mag}$ per Ce released at $T_N$ is around half of $Rln2$, and reaches $Rln2$ at around 20~K. The enhanced $C/T$ at low temperatures and reduced entropy at $T_{\rm N}$ may be due to the Kondo effect or a consequence of strongly anisotropic magnetic fluctuations above $T_N$.

\begin{figure}[h]
\includegraphics[width=8.6cm]{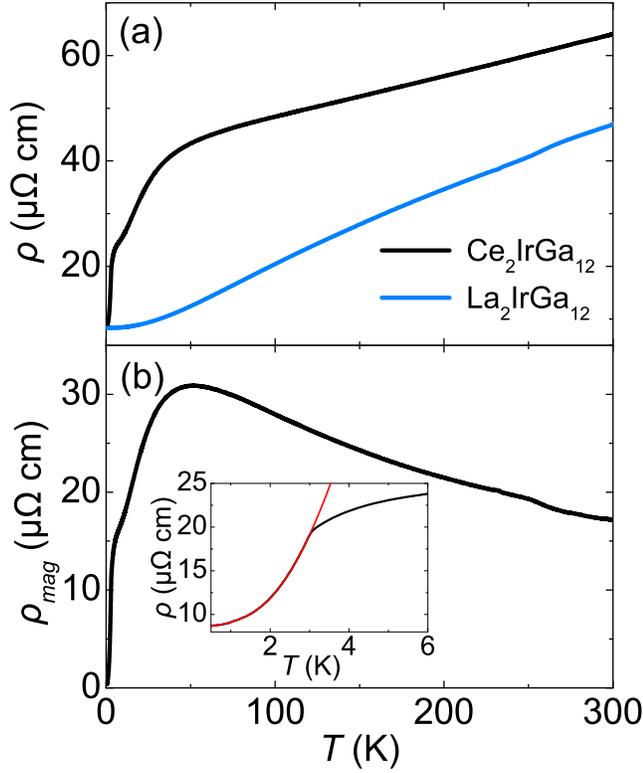}
\caption{(Color online) (a) Temperature dependence of the resistivity $\rho(T)$ of Ce$_2$IrGa$_{12}$ and La$_2$IrGa$_{12}$, (b) Temperature dependence of the magnetic contribution to the resistivity $\rho_{mag}(T)$ of Ce$_2$IrGa$_{12}$, obtained from subtracting the lattice contribution. The inset shows the low temperature $\rho(T)$ of Ce$_2$IrGa$_{12}$. The red solid line shows the results from fitting with Eq. 1.
 }
\label{fig4}
\end{figure}

\begin{figure}[h]
\includegraphics[width=8.6cm]{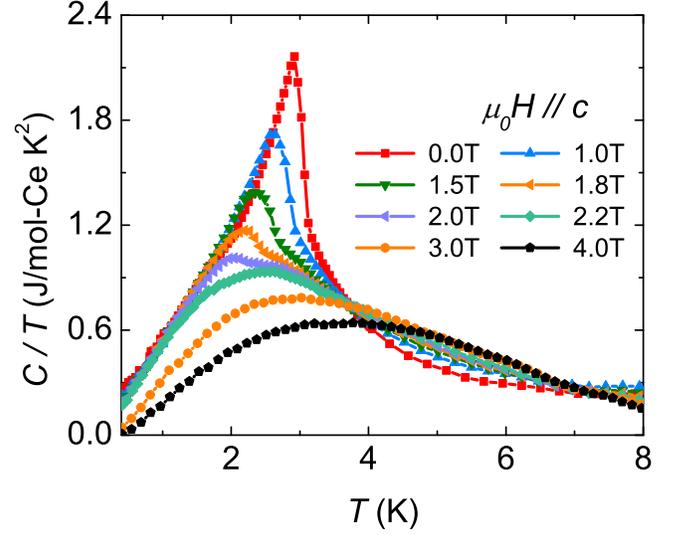}
\caption{(Color online) Temperature dependence of $C/T$ of Ce$_2$IrGa$_{12}$, which were measured in zero-field and various fields applied along the $c$~axis.
}
\label{fig5}
\end{figure}

The temperature dependence of the resistivity $\rho(T)$ of Ce$_2$IrGa$_{12}$ and nonmagnetic isostructural La$_2$IrGa$_{12}$ are shown in Fig. 4(a) down to 0.5~K, which both show metallic behavior. The residual resistivity $\rho_0$ and residual resistivity ratio RRR=$\rho(300 {\rm K})/\rho(0.5 {\rm K})$ are around 8.7~$\mu$$\Omega$~cm and 7.4 respectively, which are typical values for the samples synthesized. The magnetic contribution ($\rho_{mag}(T)$) of Ce$_2$IrGa$_{12}$, obtained from subtracting the data of La$_2$IrGa$_{12}$, are displayed in Fig. 4(b). There is a broad maximum around $T^*$=51~K in the $\rho_{mag}(T)$ of Ce$_2$IrGa$_{12}$, which is typical of Kondo-lattice systems. As shown in the inset, upon further decreasing the temperature, there is an abrupt anomaly in $\rho(T)$, corresponding to the antiferromagnetic transition. $\rho(T)$ of Ce$_2$IrGa$_{12}$ below $T_N$ can be well described by \cite{Fontes_1999}:
\begin{equation}
\rho(T)=\rho_0+AT^{2}+b\Delta^2\sqrt{\frac{T}{\Delta}}{\rm exp}(-\frac{\Delta}{T})[1+\frac{2}{3}\frac{T}{\Delta}+\frac{2}{15}(\frac{T}{\Delta})^2]
\end{equation}
where the second term corresponds to the Fermi liquid contribution, and the third term represents scattering by AFM spin-wave excitations with a spin-wave gap $\Delta$. The fitted parameters for the data in zero-field are $\rho_{0}$= 8.50(1)~$\mu\Omega$~cm, $A$=0.61(1)$\mu\Omega$~cm~K$^{-2}$, and $b$=1.64(1)$\mu\Omega$~cm~K$^{-2}$ and $\Delta$=8.53(8)~K. The sizeable value of $\Delta$ relative to $T_N$, compared to Ce$_2$PdGa$_{12}$ which has $\Delta$=16~K and $T_N$=10.5~K \cite{Gnida_2013}, suggests the presence of a stronger magnetocrystalline anisotropy in Ce$_2$IrGa$_{12}$.

\begin{figure}[t]
\includegraphics[width=8.6cm]{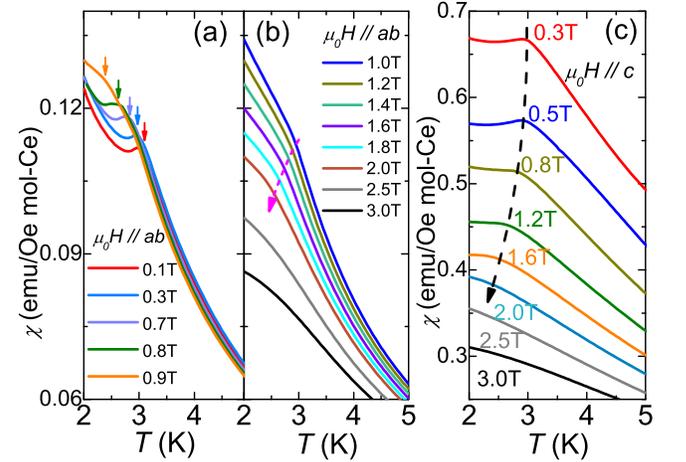}
\caption{(Color online) (a) Temperature dependence of $\chi(T)$ measured with fields applied parallel to $ab$ plane below 0.9~T. The arrows mark the antiferromagnetic transition. (b) Temperature dependence of $\chi(T)$ measured with fields applied parallel to $ab$ plane beyond 0.9~T. The dashed arrow marks the position of the anomaly observed to develop at higher fields. (c) Temperature dependence of $\chi(T)$ measured with fields applied along the $c$ axis, The dashed arrow marks the position of the antiferromagnetic transition with increasing field.
 }
\label{fig6}
\end{figure}

\begin{figure}[h]
\includegraphics[width=8.6cm]{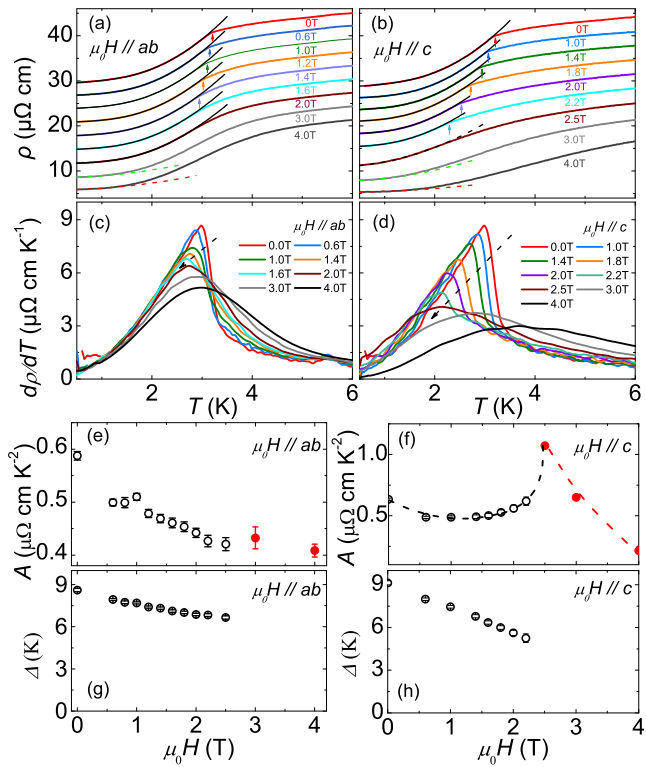}
\caption{(Color online) $\rho(T)$ of Ce$_2$IrGa$_{12}$ in various fields applied parallel to the (a) $ab$ plane and (b) $c$ axis, with the current in the $ab$ plane. The data are vertically shifted by 3~$\mu\Omega$ cm between adjacent fields, and the arrows correspond to the antiferromagnetic transition. The black solid lines show the results from fitting the data in the AFM state using Eq.1 and the dashed lines correspond to fitting with Fermi-liquid behavior at higher fields. The derivatives $d\rho(T)/dT$ are displayed for fields parallel to the (c) $ab$ plane and (d) $c$ axis. The $A$-coefficient is displayed as a function of applied field for (e) $H||ab$ and (f) $H||c$, derived from fitting the resistivity data. The black open symbols correspond to values obtained from fitting with Eq. 1 in the AFM state, while the red closed symbols are from fitting the data at higher fields using $\rho$=$\rho_0$+$AT^2$. The field-dependence of the spin wave gap $\Delta$ obtained from fitting using Eq. 1 is also displayed for (g) $H||ab$ and (h) $H||c$.
 }
\label{fig7}
\end{figure}

\subsection{Temperature-magnetic field phase diagram}

The temperature dependence of $C/T$ for Ce$_2$IrGa$_{12}$ is displayed in Fig. 5, measured with various magnetic fields applied along the $c$ axis. The jump corresponding to $T_N$ gradually shifts to lower temperature with increasing magnetic field for $\mu_0$$H<$2.2~T, as expected for an antiferromagnetic transition. For fields higher than 2.2~T, no transition is observed but there is a broad maximum, which moves to higher temperatures with increasing field. This likely corresponds to a Schottky contribution arising from Zeeman splitting of the ground-state doublet. No divergence is observed in $C/T$ upon suppressing the magnetism with applied magnetic field, indicating a lack of field induced quantum criticality in Ce$_2$IrGa$_{12}$.

The temperature dependence of $\chi(T)$ in various fields for two orientations is displayed in Figs. 6(a) and Figs. 6(b) for $H||ab$, and Figs. 6(c) for $H||c$. In a field of 0.1~T, $\chi(T)$ shows a peak at 3.1~K both for $H||ab$ and $H||c$, corresponding to the antiferromagnetic transition. For $H||ab$, the peak positions marked by the arrows in Fig. 6(a), are suppressed rapidly to lower temperatures with increasing magnetic field, disappearing above 0.9~T. Above 0.9~T, a weaker anomaly is observed to emerge at higher temperatures, as indicated by the dashed line in Fig. 6(b). Upon further increasing field, this feature broadens and gradually moves to lower temperatures, and is difficult to detect above 2~T. Meanwhile for $H||c$, as shown in Fig. 6(c), a clear anomaly is observed at $T_N$ below which $\chi(T)$ is nearly temperature independent. This transition is suppressed with field, and is not observed in a 2~T field.

$\rho(T)$ of Ce$_2$IrGa$_{12}$ for various fields applied parallel to the $ab$ plane and $c$ axis are shown in Figs. 7(a) and 7(b) respectively, with the current in the $ab$ plane. The corresponding derivatives $d\rho(T)/dT$ are displayed in Figs. 7(c) and (d). At low fields, an anomaly is observed in $\rho(T)$ corresponding to the antiferromagnetic transition, and there is a sharp asymmetric peak in $d\rho(T)/dT$. At higher fields, a clear anomaly is not observed in $\rho(T)$, and $d\rho(T)/dT$ instead exhibits a broad symmetric peak, indicating the lack of a magnetic transition. For $H||ab$, the transition is suppressed by field, but for fields below 0.9 T, the suppression appears to be less rapid with field than that of the peak in $\chi(T)$ (Fig. 6(a)). Meanwhile the anomaly becomes less pronounced with increasing applied field, and is difficult to detect beyond 1.4~T.  For $H||c$, $T_N$ is also gradually suppressed to lower temperatures with increasing magnetic field, and the magnetic transitions are not observed at fields beyond 2.2~T.  For both $H||ab$ and $H||c$, $\rho(T)$ follows Fermi liquid behavior at higher fields, as shown by the dashed lines.

The in-field $\rho(T)$ data in the AFM state for both field directions were also analyzed using Eq. 1, and the field dependence of the $A$-coefficient and $\Delta$ for $H||ab$ and $H||c$ are displayed in Figs. 7(e)-(h). $\rho(T)$ were also analyzed at higher fields in the spin-polarized state using the Fermi liquid expression $\rho(T)$=$\rho_0$ + $AT^2$. For $H||ab$, the $A$-coefficient exhibits a drop upon the application of a field, which may be related to the emergence of the field induced AF2 phase (see below). Upon further increasing the field, $A$ continues to decrease, showing no enhancement at the transition to the spin-polarized phase. On the other hand, for $H||c$ the $A$-coefficient shows an abrupt increase at the transition from the AFM to spin-polarized phase, reaching a maximum value at 2.5~T, before decreasing at higher fields. Such a pronounced enhancement of the $A$-coefficient at the field-induced metamagnetic transition to the spin-polarized phase is very similar to  that observed in the heavy fermion Ising antiferromagnet CeRh$_2$Si$_2$ for fields along the easy c-axis \cite{Knafo_2010, Knafo_2017}. Here this was ascribed to a significant enhancement of the effective mass, driven by the presence of critical magnetic fluctuations. Moreover, a similar enhancement is also observed in CeRh$_2$Si$_2$ at the pressure induced suppression of antiferromagnetic order, in the vicinity of the superconducting dome \cite{Araki_2002}. The field dependence of $\Delta$ for $H||ab$ and $H||c$ is displayed in Figs. 7(g) and (h). The magnetic field reduces $\Delta$ for both field directions, which is the expected behavior for antiferromagnetic systems \cite{Jobiliong_2005}.

 Figures 8(a) and (b) display the isothermal magnetization $M(H)$ of Ce$_2$IrGa$_{12}$ at 2.0~K for fields in the $ab$-plane and along the $c$-axis. As displayed in Fig. 8(a), the magnetization for $H||ab$ at 2.0~K exhibits a kink at $B_m$=0.9~T and a broad shoulder at $B_P$=2.5~T. From comparing the $\chi(T)$ and $\rho(T)$ data, $B_m$ likely corresponds to a metamagnetic transition from the low field antiferromagnetic phase to a field induced magnetic state, while the broad shoulder at $B_P$=2.5~T may be the crossover from this field-induced phase to the spin-polarized state. In Fig. 8(b), the isothermal magnetization indicates that the $c$-axis is the easy axis of magnetization, reaching 1.64~$\mu_B$/Ce for $H||c$ at 9~T and 2~K, compared to 0.58~$\mu_B$/Ce for $H||ab$. For $H||c$, the magnetization at 2~K only shows a broad crossover at 2.3~T, above which there is a much less rapid change with field, again suggesting a change to the spin polarized state. There is a lack of hysteresis in both $M(H)$ and $\rho(H)$ upon increasing and decreasing the field, suggesting a second-order nature of the transitions.

\begin{figure}[h]
\includegraphics[width=8.6cm]{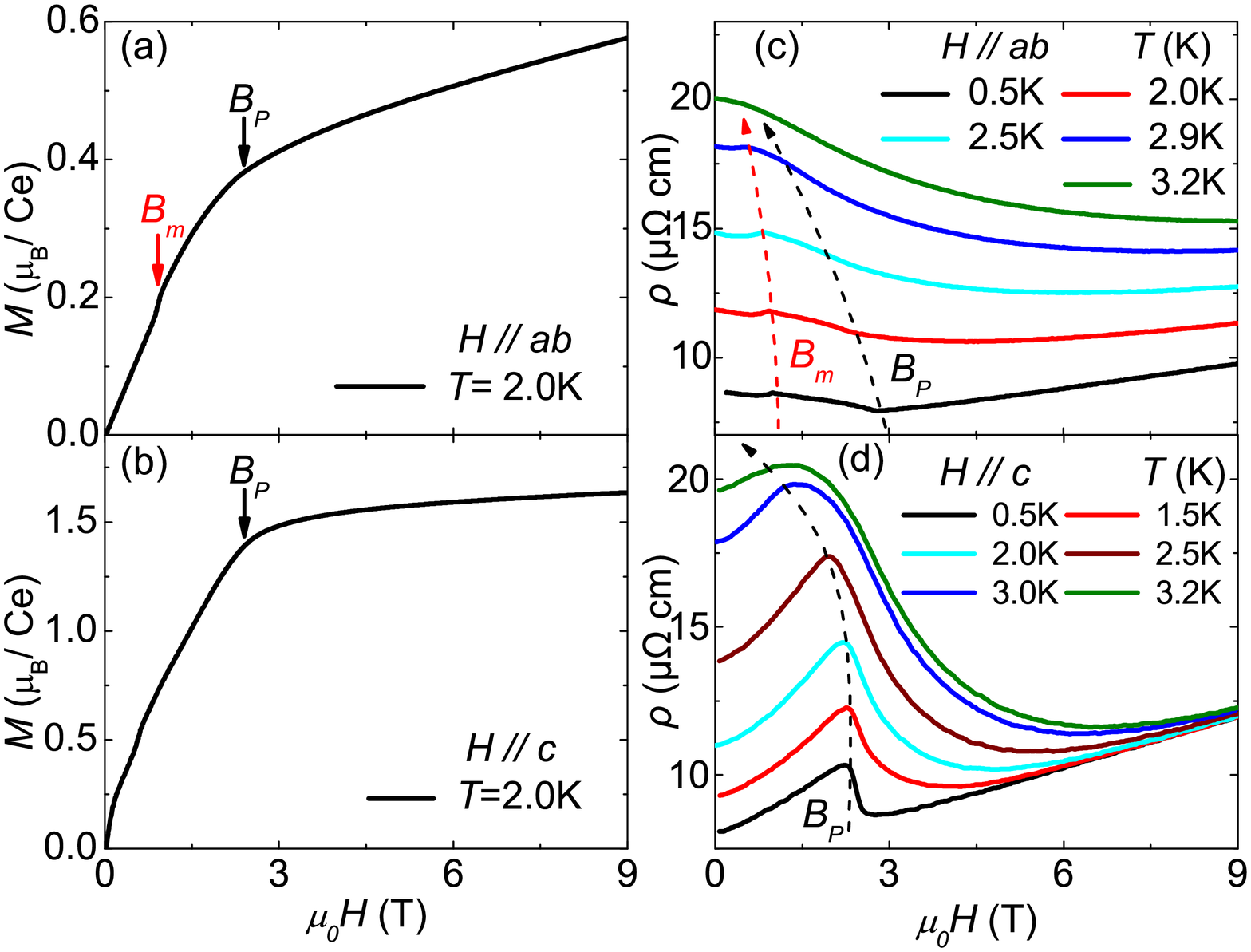}
\caption{(Color online) Isothermal magnetization $M(H)$ of Ce$_2$IrGa$_{12}$ as a function of field at 2.0~K for (a) $H||ab$ and (b) $H||c$. Vertical arrows mark the metamagnetic transition $B_m$ and the crossover $B_p$. The isothermal field dependence of the resistivity $\rho(H)$ of Ce$_2$IrGa$_{12}$ is displayed at various temperatures for fields applied parallel to the (c) $ab$ plane and (d) $c$ axis. The dashed lines mark the trend of $B_m$ and $B_p$ with temperature.
 }
\label{fig8}
\end{figure}

\begin{figure}[h]
\includegraphics[width=8.6cm]{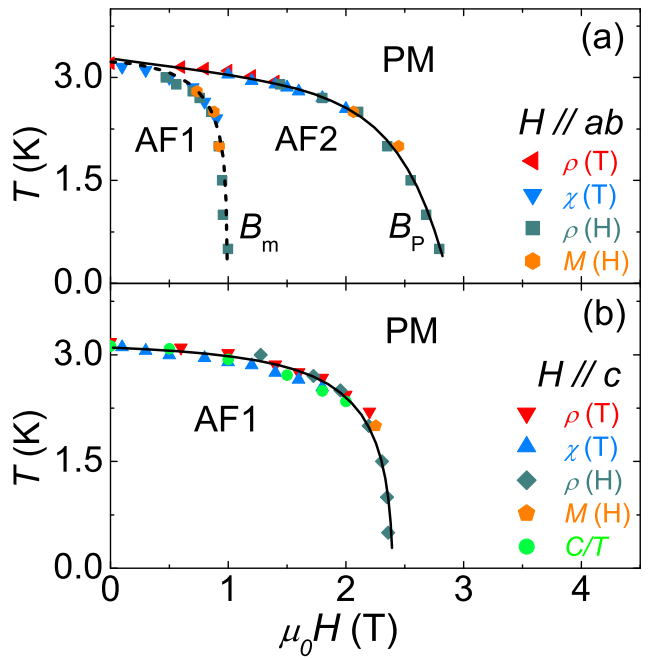}
\caption{(Color online)
The temperature-field phase diagram of Ce$_2$IrGa$_{12}$ for magnetic fields applied (a) within the $ab$-plane, and (b) along the $c$ axis, based on specific heat, resistivity, magnetoresistivity and magnetization measurements.
 }
\label{fig9}
\end{figure}

\begin{figure}[h]
\includegraphics[width=8.6cm]{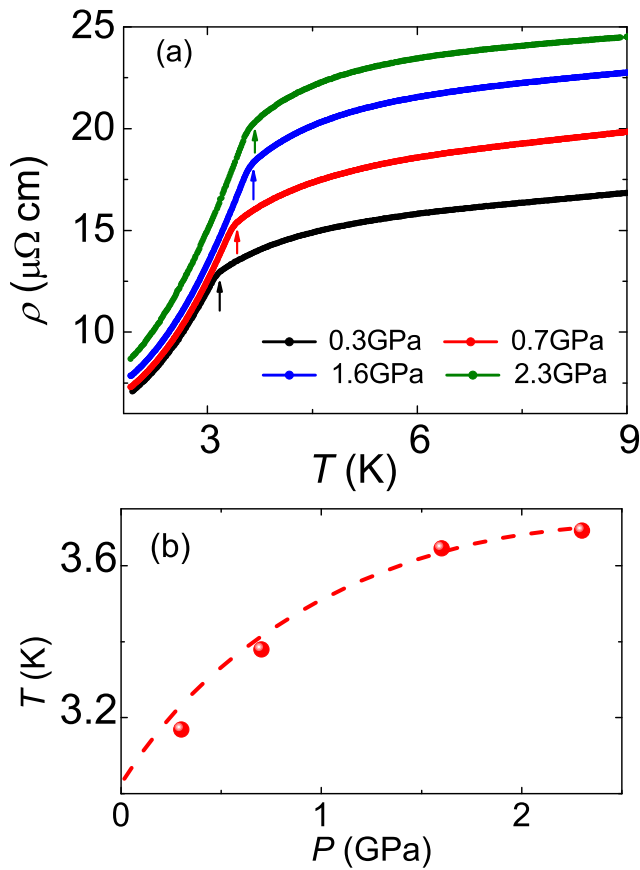}
\caption{(Color online) (a) $\rho(T)$ of Ce$_2$IrGa$_{12}$ under various various pressures, in zero-applied field,  the arrows correspond to the transition at $T_N$. (b) $T-P$ phase diagram of Ce$_2$IrGa$_{12}$, where the symbols correspond to $T_N$, and the dashed line is a guide to the eye.
}
\label{fig10}
\end{figure}

The field-dependent resistivity $\rho(H)$ of Ce$_2$IrGa$_{12}$ at different temperatures is displayed in Fig. 8(c) for $H||ab$ and Fig. 8(d) for $H||c$. For $H||ab$, $\rho(H)$ at 0.5~K exhibits a peak at 1.0~T and a clear minimum with a sharp cusp at 2.8~T, which correspond to the metamagnetic transition at $B_m$ and crossover at $B_P$ respectively. Upon increasing the temperature, both $B_m$ and $B_P$ shift to lower fields, as indicated by the dashed lines, before disappearing above $T_N$. In the case of $H||c$, a broad hump is observed at 2.4~T at 0.5~K, which shifts to lower field with increasing temperature, indicating a crossover to the spin polarized state.

Based on the measurements of resistivity, susceptibility and specific heat of Ce$_2$IrGa$_{12}$ with magnetic fields applied in the $ab$ plane and along the $c$ axis, we constructed the $T-H$ phase diagrams, which are displayed in Fig. 9. For $H||ab$, the two boundaries of $B_m$ and $B_P$ are also shown in the phase diagram, and for in-plane fields there exists both a low-field antiferromagnetic phase (AF1), and a field-induced phase (AF2). $B_m$ corresponds to the metamagnetic transition between the AF1 and AF2 phases, while $B_P$ separates the AF2 and spin-polarized phases. For $H||c$, the phase boundaries obtained from $C(T)$, $\rho(T)$, $\rho(H)$ and $\chi(T)$ are all consistent, where $T_N$ is continuously suppressed with field, before being no longer observed beyond 2.4~T.

\subsection{Phase diagram in Ce$_2$IrGa$_{12}$ under pressure }

In order to examine the effects of hydrostatic pressure on the magnetic order and to look for  pressure induced quantum criticality in Ce$_2$IrGa$_{12}$, $\rho(T)$ was measured under various hydrostatic pressures up to 2.3~GPa, as displayed in Fig. 10(a). With increasing pressure, the magnitude of $\rho(T)$ above $T_N$ increases, which may be due to enhanced Kondo scattering or magnetic scattering. $T_N$ was determined from the onset of the resistivity anomaly, and the resulting phase diagram is displayed in Fig. 10(b). The $T_N$ of Ce$_2$IrGa$_{12}$ is found to be slightly enhanced upon applying pressure, to around 3.7~K at 2.3~GPa.

\section{DISCUSSION}

Ce$_2$IrGa$_{12}$ crystallizes in the tetragonal space group, $P4/nbm$ (No.125), in which the layers of CeGa$_6$ are inserted between two layers of IrGa$_6$, analogous to the structure of Ce$_n$$M$In$_{3n+2}$. In Ce$_2$IrGa$_{12}$, the Ce-Ce interatomic distances are 3.031~\AA~in the $ab$ plane and 8.073~\AA~along the c-axis, This suggests the presence of anisotropic magnetic exchange interactions, which is supported by the anisotropic Curie-Weiss temperatures of $\theta_P$=-22.6~K ($H||ab$) and 10.2~K ($H||c$). Anisotropic magnetic behaviors are also detected in other Ce$_2M$Ga$_{12}$ compounds, such as Ce$_2$RhGa$_{12}$, which also has an easy $c$-axis and $\theta_P$  of -39~K for $H||ab$ and 5~K for $H||c$ \cite{Nallamuthu_2014}.

The $T_N$ values of Ce$_2$$M$Ga$_{12}$ for $M$=Pd, Ni, Rh and Ir are 11~K, 10~K, 3.5~K and 3.1~K, respectively \cite{ Macaluso_2005, Cho_2008, Nallamuthu_2014}. The unit cell volume of Ce$_2$IrGa$_{12}$ (576.16~\AA$^3$) is slightly larger than that of Ce$_2$NiGa$_{12}$ (564.9~\AA$^3$), while it is much smaller than that of Ce$_2$PdGa$_{12}$ (579.64~\AA$^3$) \cite{ Macaluso_2005, Cho_2008}. However, the $T_N$ of Ce$_2$IrGa$_{12}$ is much lower than both the aforementioned compounds, suggesting that the substition of different transition metals does not simply correspond to a chemical pressure effect, but may otherwise tune the system, such as by modifying the electronic structure or carrier density. On the other hand, the substitution of Ir with Rh does correspond to a positive chemical pressure, since the lattice volume of Ce$_2$RhGa$_{12}$ is 573.1~\AA$^3$, and $T_N$ is enhanced slightly \cite{Nagalakshmi_2011,Nallamuthu_2014}, which is consistent with our measurements of Ce$_2$IrGa$_{12}$ under pressure. Furthermore the $\gamma$ obtained from the $C/T$ data above $T_N$ is 212~mJ/mol-Ce~K$^2$ in Ce$_2$RhGa$_{12}$ \cite{Nallamuthu_2014}, which is larger than the value of 144.4~mJ/mol-Ce~K$^2$ we obtain for Ce$_2$IrGa$_{12}$.

The $T_N$ of Ce$_2$IrGa$_{12}$ is enhanced to 3.7~K, upon applying a pressure of 2.3~GPa. The moderate increase of $T_N$ indicates that Ce$_2$IrGa$_{12}$ is situated on the left side of Doniach phase diagram, with weak coupling and well localized $4f$ electrons \cite{Doniach_1977}. Therefore, to look for quantum criticality in Ce$_2$IrGa$_{12}$, significantly larger pressures are likely necessary, in order to sufficiently enhance the relative strength of the Kondo interaction. From Fig. 10(b), the $T_N$ at 2.3~GPa of 3.7~K appears to be close to the maximum value, which is very similar to the $T_N$ of Ce$_2$RhGa$_{12}$. On the other hand, Ce$_2$NiGa$_{12}$ and Ce$_2$PdGa$_{12}$ have considerably larger ordering temperatures, which are suppressed with pressure \cite{Ohara_2012,Kawamura_2014}. These results suggest that in the Ce$_2$$M$Ga$_{12}$ family, the systems with $M$=Pd and Ni have larger energy scales for the magnetism than those with $M$=Rh and Ir.

\section{Conclusion}

In summary, we have successfully synthesized single crystals of Ce$_2$IrGa$_{12}$ using a flux method, which crystallizes in a layered tetragonal structure with space group $P4/nbm$ (No. 125). We find that Ce$_2$IrGa$_{12}$ orders antiferromagnetically below $T_N$=3.1K. We construct the temperature-field phase diagrams for both fields within the $ab$-plane and along the $c$-axis, where we find evidence for a metamagnetic transition to a different field-induced phase for in-plane fields. The reduced entropy at $T_N$ and enhanced low temperature $C/T$ may  be a consequence of the Kondo effect, or short range magnetic fluctuations, where the presence of the latter is also supported by a strongly anisotropic negative magnetoresistivity above $T_N$. Upon applying pressure, $T_N$ undergoes a moderate enhancement for pressures up to 2.3~GPa. As a result, measurements at higher pressures are necessary to examine for the presence of quantum criticality, while neutron diffraction experiments could reveal the magnetic structure of both the zero-field and field-induced phases.

\section{acknowledgments}

We are grateful to X. Lu, F. Steglich and KiSoo Park for interesting discussions and helpful suggestions. This work was supported by the National Key R\&D Program of China (No.~2017YFA0303100, No.~2016YFA0300202 and No.~2016YFA0401704), the Users with Excellence Project of Hefei Science Center CAS (Grant No. 2018HSC-UE012), the National Natural Science Foundation of China (No.~U1632275, No.~11604291), the Science Challenge Project of China (No.~TZ2016004), the Korea Research Foundation(KRF) Grants No. 2018R1D1A1B07049479, and the Basic Science Research Program through the National Research Foundation of Korea (NRF) funded by the Ministry of Education (NRF-2018R1D1A1B07049479).

\end{document}